\documentstyle[12pt]{article}

\newcommand{\cl}{\centerline}
\renewcommand{\theequation}{\arabic{equation}}
\newcommand\beq{\begin{equation}}
\newcommand\eeq{\end{equation}}
\newcommand\bea{\begin{eqnarray}}
\newcommand\eea{\end{eqnarray}}

\begin{document}

\begin{titlepage}
\setlength{\textwidth}{5.0in}
\setlength{\textheight}{7.5in}
\setlength{\parskip}{0.0in}
\setlength{\baselineskip}{18.2pt}
\begin{center}
{\Large\bf Symmetries of SU(2) Skyrmion in Hamiltonian and Lagrangian approaches}
\par
\begin{center}
{Soon-Tae Hong, Yong-Wan Kim and Young-Jai Park}\par
\end{center}
\begin{center}
{Department of Physics and Basic Science Research Institute,}\par
{Sogang University, C.P.O. Box 1142, Seoul 100-611, Korea}\par
\end{center}
\cl{\today}
\vfill
\begin{center}
{\bf ABSTRACT}
\end{center}
\begin{quotation}
We apply the Batalin-Fradkin-Tyutin (BFT) method to the SU(2) Skyrmion to
study the full symmetry structure of the model at the first class Hamiltonian level.
On the other hand, we also analyze the symmetry structure of the action
having the WZ term, which corresponds to this Hamiltonian,
in the framework of the Lagrangian approach.
Furthermore, following the BFV formalism we derive
the BRST invariant gauge fixed Lagrangian from
the above extended action.

\noindent
\end{quotation}
\end{center}
\end{titlepage}


\section{Introduction}

\setcounter{equation}{0} \renewcommand{\theequation}{\arabic{section}.%
\arabic{equation}}

It is well known that baryons can be obtained from topological solutions,
known as SU(2) Skyrmions, since the homotopy group $\Pi_{3}(SU(2))=Z$ admits
fermions \cite{ad,hsk}. Using the collective coordinates of the isospin
rotation of the Skyrmion, Adkins et al. \cite{ad} have performed
semiclassical quantization to obtain the static properties of baryons within
about 30$\%$ of the corresponding experimental data.

On the other hand, in order to quantize physical systems subjective to
constraints, Dirac quantization scheme \cite{di} has been widely used.
However, whenever we adopt the Dirac method, we frequently meet the problem
of operator ordering ambiguity. In order to avoid this problem, Batalin,
Fradkin, and Tyutin (BFT) developed a method \cite{BFT}, which converts
second class constraints into first class ones by introducing auxiliary
fields. Recently, we have clarified the relation
between the Dirac scheme and BFT one, which has been obscure and
unsettled up to now, in the framework of the SU(2) Skyrmion model \cite{sk2}.

The motivation of this paper is to systematically apply the BFT and
Batalin, Fradkin, and Vilkovisky (BFV)-BRST method to the SU(2) Skyrmion
as a phenomenological example of topological system,
and to consider the problem of finding all local symmetries
of the system through both the Hamiltonian and Lagrangian approaches.
As a result, we will show that these approaches give
the same symmetry structure of the SU(2) Skyrmion.
In section 2, we will briefly recapitulate the construction of
the first class SU(2) Skyrmion Hamiltonian.
In section 3, we will study full symmetry structure of the system in the
pure Hamiltonian approach recently proposed \cite{brr}.
In section 4, we will treat symmetry structure of the Lagrangian,
which includes the Wess-Zumino (WZ) term, corresponding to
the first class Hamiltonian in the Lagrangian approach,
to compare with the results in the previous section.
Finally, we will construct the
BRST invariant gauge fixed Lagrangian from the extended one
corresponding to the first class Hamiltonian in the
BFV scheme \cite{bfv,fik,biz,kpy}.


\section{BFT Hamiltonian for SU(2) Skyrmion}

\setcounter{equation}{0} \renewcommand{\theequation}{\arabic{section}.%
\arabic{equation}}


Now we start with the Skyrmion Lagrangian of the form
\begin{equation}
L_{SM}=\int{\rm d}r^{3}[- \frac{f_{\pi}^{2}}{4}{\rm tr}(\partial_{\mu}U^{%
\dag}\partial^{\mu}U)+\frac{1} {32e^{2}}{\rm tr}[U^{\dag}\partial_{\mu}U,U^{%
\dag}\partial_{\nu}U]^{2}]
\end{equation}
where $f_{\pi}$ is the pion decay constant and $e$ is a dimensionless
parameter and $U$ is an SU(2) matrix satisfying the boundary condition $%
\lim_{r \rightarrow \infty} U=I$ so that the pion field vanishes as $r$ goes
to infinity.  On the other hand, in the Skyrmion model, spin and isospin states
can be treated by collective coordinates $a^{\mu}=(a^{0},\vec{a})$ $(\mu=0,1,2,3)$
corresponding to the spin and isospin rotations
\begin{equation}
A(t) = a^{0}+i\vec{a}\cdot\vec{\tau}.
\end{equation}
With the hedgehog ansatz and the collective rotation $A(t)\in$ SU(2), the Skyrmion
Lagrangian can be written as
\begin{equation}
L_{SM}=-E+2{\cal I}\dot{a}^{\mu}\dot{a}^{\mu}   \label{lag}
\end{equation}
where $E$ and ${\cal I}$ are the soliton energy and the moment of inertia,
respectively\cite{ad,sk2}.  Introducing the canonical momenta $\pi^{\mu}=4{\cal I}\dot{a}^{\mu}$ conjugate to
the collective coordinates $a^{\mu}$ one can then obtain the canonical Hamiltonian
\begin{equation}
H=E+\frac{1}{8{\cal I}}\pi^{\mu}\pi^{\mu}.  \label{hamil}
\end{equation}
Then, our system has the following second class constraints
\begin{eqnarray}
\Omega_{1} &=& a^{\mu}a^{\mu}-1\approx 0,  \nonumber \\
\Omega_{2} &=& a^{\mu}\pi^{\mu}\approx 0 ,  \label{omega2}
\end{eqnarray}
to yield the Poisson algebra with $\epsilon^{12}=-\epsilon^{21}=1$
\begin{equation}
\Delta_{k k^{\prime}}=\{\Omega_{k},\Omega_{k^{\prime}}\} = 2\epsilon^{k
k^{\prime}}a^{\mu}a^{\mu}  \label{delta}.
\end{equation}

Next, let us briefly recapitulate the construction of the first
class SU(2) Hamiltonian.
Following the abelian BFT formalism \cite{BFT,sk2} which
systematically converts the second class constraints into first class ones,
we introduce two auxiliary fields $\Phi^{i}$ corresponding to $\Omega_{i}$
with the Poisson brackets
\begin{equation}
\{\Phi^{i}, \Phi^{j}\}=\epsilon^{ij}.  \label{phii}
\end{equation}
One can then obtain the following first class constraints
\begin{eqnarray}
\tilde{\Omega}_{1}&=&\Omega_{1}+2\Phi^{1}  \nonumber \\
\tilde{\Omega}_{2}&=&\Omega_{2}-a^{\mu}a^{\mu}\Phi^{2}  \label{1stconst}
\end{eqnarray}
satisfying the first class constraint algebra $\{\tilde{\Omega}_{i},\tilde{\Omega}_{j}\}=0$.
Then, by demanding that they are strongly involutive in the extended phase space, i.e.,
$\{\tilde{\Omega}_{i}, \tilde{{\cal F}}\}=0$, we construct the first class BFT
physical fields $\tilde{{\cal F}}=(\tilde{a}^{\mu}, \tilde{\pi}^{\mu})$ corresponding
to the original fields ${\cal F}=(a^{\mu},\pi^{\mu})$, as a power series of the auxiliary
fields $\Phi^{i}$, as follows\footnote{
Here one notes that the Poisson brackets of $\tilde{{\cal F}}$'s have the same
structure as that of the corresponding Dirac brackets \cite{sk2}.}
\begin{eqnarray}
\tilde{a}^{\mu}&=&a^{\mu}\left[1-\sum_{n=1}^{\infty}\frac{(-1)^{n}(2n-3)!!}{n!}
\frac{(\Phi^{1})^{n}}{(a^{\mu}a^{\mu})^{n}}\right]  \nonumber \\
\tilde{\pi}^{\mu}&=&(\pi^{\mu}-a^{\mu}\Phi^{2})\left[1+\sum_{n=1}^{\infty}
\frac{(-1)^{n}(2n-1)!!}{n!}\frac{(\Phi^{1})^{n}}{(a^{\mu}a^{\mu})^{n}}\right].
\label{pitilde}
\end{eqnarray}

Now, exploiting the novel property \cite{kpy,kpr} that any functional ${\cal K}(%
\tilde{{\cal F}})$ of the first class fields $\tilde{{\cal F}}$ will also be
first class, i.e.,
$$
\tilde{{\cal K}}({\cal F};\Phi )={\cal K}(\tilde{{\cal F}})  \label{ktilde}
$$
one can directly construct the first class Hamiltonian in terms of the above
BFT physical variables as follows
\begin{equation}
\tilde{H}=E+\frac{1}{8{\cal I}}\tilde{\pi}^{\mu}\tilde{\pi}^{\mu}
\label{htilde}
\end{equation}
omitting infinitely iterated standard procedure \cite{sk2}. As a result,
the corresponding first class Hamiltonian with the original fields and
auxiliary fields is given by
\begin{equation}
\tilde{H}=E+\frac{1}{8{\cal I}}(\pi^{\mu}-a^{\mu}\Phi^{2})
(\pi^{\mu}-a^{\mu}\Phi^{2})\frac{a^{\nu}a^{\nu}}{a^{\nu}a^{\nu}+2\Phi^{1}}
\label{hct}
\end{equation}
which is also strongly involutive with the first class constraints
\begin{equation}
\{\tilde{\Omega}_{i},\tilde{H}\}=0.
\end{equation}
However, with the first class Hamiltonian (\ref{hct}), one cannot naturally
generate the first class Gauss' law constraint from the time evolution of
the primary constraint $\tilde{\Omega}_{1}$. Now, by introducing an additional term
proportional to the first class constraints $\tilde{\Omega}_{2}$ into $%
\tilde{H}$, we obtain an equivalent first class Hamiltonian
\begin{equation}
\tilde{H}^{\prime}=\tilde{H}+\frac{1}{4{\cal I}}\Phi^{2} \tilde{\Omega}_{2}
\label{hctp}
\end{equation}
which naturally generates the Gauss' law constraint
\begin{equation}
\{\tilde{\Omega}_{1},\tilde{H}^{\prime}\}=\frac{1}{2{\cal I}}
\tilde{\Omega}_{2},~~~
\{\tilde{\Omega}_{2},\tilde{H}^{\prime}\}=0. \label{ga}
\end{equation}
Here one notes that $\tilde{H}$ and $\tilde{H}^{\prime}$ act on physical
states in the same way since such states are annihilated by the first class
constraints.

It seems appropriate to comment on the phenomenolocal application of
the above Hamiltonian $\tilde{H}^{\prime}$.
Using the first class constraints in this Hamiltonian (\ref{hctp}), one
can finally obtain the Hamiltonian of the form
\begin{equation}
\tilde{H}^{\prime}=M+\frac{1}{8{\cal I}}(a^{\mu}a^{\mu}\pi^{\nu}\pi^{\nu}
-a^{\mu}\pi^{\mu}a^{\nu}\pi^{\nu}).  \label{htilde2}
\end{equation}
Following the symmetrization procedure, the first class Hamiltonian yields
the slightly modified energy spectrum with the Weyl ordering correction \cite{sk2}
\begin{equation}
\langle\tilde{H}^{\prime}\rangle=E+\frac{1}{2{\cal I}}[I(I+1)+\frac{1}{4}]
\label{nht}
\end{equation}
where $I$ is the isospin quantum number of baryons.


\section{Symmetry Structure of First Class Hamiltonian}
\setcounter{equation}{0} \renewcommand{\theequation}{\arabic{section}.%
\arabic{equation}}


Now, since we have successfully converted the second class SU(2) Skyrmion
into the corresponding first class one with the BFT scheme,
we are ready to unravel gauge symmetries of the first
class system following the purely Hamiltonian approach \cite{brr},
which is very recently proposed.
This total Hamiltonian approach\footnote{For an extended Hamiltonian approach,
see the work of \cite{htz}. This extended Hamiltonian approach with
suitable gauge conditions is equivalent to the total Hamiltonian approach.
See also Ref. \cite{brr}.}
is based on the requirement of the commutativity
of a general gauge variation with the time derivative operation
which puts restrictions on gauge parameters and Lagrange multipliers.

Following Dirac's conjecture \cite{di}, let us first construct the generators
of gauge transformation for the SU(2) Skyrmion model which has totally two
constraints as
\begin{equation}
G=\epsilon^a \tilde{\Omega}_a, ~~~~a=1,2.
\end{equation}
Here, $\epsilon^a$ are in general functions of phase space variables and
$\tilde{\Omega}_a$ are first class constraints in Eq.$(\ref{1stconst})$. Then,
the infinitesimal gauge transformation is given by the relation of
$\delta F(p,q) = \{ F(p,q), G \}$ in which $F$ is a function of phase space
variables.
The total Hamiltonian is easily read as
\begin{equation}
\tilde{H}_{\rm T} = \tilde{H}' + \lambda \tilde{\Omega}_1,
\end{equation}
where $\tilde{H}'$ is the canonical Hamiltonian $\tilde{H}_{\rm C}$
and $\tilde{\Omega}_1$ the primary first class constraint.
Comparing the general expression of gauge algebra
\begin{eqnarray}
\{\tilde{\Omega}_a, \tilde{H}_{\rm C}\} &=& V_a^b \tilde{\Omega}_b, \nonumber \\
\{\tilde{\Omega}_a, \tilde{\Omega}_b\} &=& C^{c}_{ab} \tilde{\Omega}_c
\end{eqnarray}
with Eq. (\ref{ga}), we can determine the gauge functions of $V_a^b$ as well
as $C^{c}_{ab}$. Note that the subscripts $a$ and $b$ count all the number
of constraints, while denote the subscript of the primary constraints as $a_1$
and the secondary constraints as $a_2$.

Now the requirement of the commutativity of the general variation with the
time derivative operator gives the relation \cite{brr} as follows
\begin{eqnarray}
\delta v^{b_1} &=& \frac{d\epsilon^{b_1}}{dt}
               -\epsilon^a(V^{b_1}_a+v^{a_1}C^{b_1}_{a_1a}), \label{first1} \\
0&=& \frac{d\epsilon^{b_2}}{dt}
               -\epsilon^a(V^{b_2}_a+v^{a_1}C^{b_2}_{a_1a}). \label{second2}
\end{eqnarray}
Here, $v^{b_1}$ are Lagrange multipliers associated with the primary first
class constraints in the total Hamiltonian, and Eq.(\ref{second2}) gives
the restriction on the gauge parameters.

Since there exist one primary and one secondary constraints for our
SU(2) Skyrmion case, we easily see that the condition (\ref{second2}) imposed
on the gauge parameters is simply rewritten as the relation of $\dot{\epsilon}^2=
\frac{1}{2{\cal I}}\epsilon^1$. Moreover, making use of this relation,
we explicitly obtain the infinitesimal gauge transformation of
the field variables as follows
\begin{equation}
\delta a = \{a^\mu, G\}= a^\mu \epsilon,
~~~\delta \Phi^1 =\{\Phi^1,G\}=-a^\mu a^\mu \epsilon, \label{rule}
\end{equation}
where we have rewritten the independent gauge paramether $\epsilon^2$
as $\epsilon$.

As a result of applying the approach on a purely Hamiltonian level to
the SU(2) Skyrmion model, we have derived the rule of
the full symmetry transformation.


\section{Symmetry Structure of Corresponding Lagrangian}
\setcounter{equation}{0} \renewcommand{\theequation}{\arabic{section}.%
\arabic{equation}}


Now let us consider the partition function of the model in order to present
the Lagrangian corresponding to the first class Hamiltonian $\tilde{H}%
^{\prime}$ in Eq.(\ref{hctp}). First of all we identify
the auxiliary fields $\Phi^{i}$ with a canonical conjugate pair $%
(\theta,\pi_{\theta})$, i.e.,
\begin{equation}
\Phi^{i}=(\theta,\pi_{\theta})
\end{equation}
which satisfy Eq. (\ref{phii}). Then, the starting
partition function in the phase space is given by
the Faddeev-Senjanovic formula \cite{faddev} as follows
\begin{eqnarray}
Z&=&N\int {\cal D}a^{\mu}{\cal D}\pi^{\mu}{\cal D}\theta{\cal D}\pi_{\theta}
\prod_{i,j=1}^{2}\delta(\tilde{\Omega}_i)\delta(\Gamma_j)\det|\{\tilde
{\Omega}_i,\Gamma_j\}|e^{i\int {\rm d}t L}  \nonumber \\
L&=&\dot{a}^{\mu}\pi^{\mu}+\dot{\theta}\pi_\theta-\tilde{H}^{\prime}
\end{eqnarray}
where the gauge fixing conditions $\Gamma_{i}$ are chosen so that the
determinant occurring in the functional measure is nonvanishing.

Now, exponentiating the delta function $\delta (\tilde{\Omega}_{2})$ as
$\delta (\tilde{\Omega}_{2})=\int {\cal D}\xi e^{i\int {\rm d}t~\xi
\tilde{\Omega}_{2}}$ and performing the integration over $\pi_{\theta}$,
$\pi^{\mu}$ and $\xi$, we obtain the the following partition function
\begin{eqnarray}
Z&=&N\int {\cal D}a^{\mu}{\cal D}\theta\delta(a^{\mu}a^{\mu}-1+2\theta)
\prod_{i=1}^{2}\delta(\Gamma_i)\det|\{\tilde{\Omega}_i,\Gamma_j\}| e^{i\int
{\rm d}t L}  \label{fca} \\
L&=& -E+\frac{2{\cal I}}{a^{\sigma}a^{\sigma}} \dot{a}^{\mu}\dot{a}^{\mu}-%
\frac{2{\cal I}}{(a^{\sigma}a^{\sigma})^{2}} \dot{\theta}^{2}.  \label{zhct}
\end{eqnarray}
As a result, we have obtained the desired Lagrangian (\ref{zhct})
corresponding to the first class Hamiltonian (\ref{hctp}).
Here one notes that the Lagrangian (\ref{zhct}) can be reshuffled to yield
the gauge invariant action of the form
\begin{eqnarray}
S&=& \int {\rm d}t \ (-E+2{\cal I}\dot{a}^\mu\dot{a}^\mu ) +S_{WZ}  \nonumber
\\
S_{WZ}&=&\int {\rm d}t \left[ \frac{4{\cal I}}{a^{\sigma}a^{\sigma}}
\dot{a}^{\mu}\dot{a}^{\mu}\theta
-\frac{2{\cal I}}{(a^{\sigma}a^{\sigma})^{2}} \dot{%
\theta}^{2} \right],
\end{eqnarray}
where $S_{WZ}$ is the new type of the Wess-Zumino term restoring the gauge
symmetry. Moreover the corresponding partition function (\ref{fca}) can be
rewritten simply in terms of the first class physical fields (\ref{pitilde})
\begin{eqnarray}
\tilde{Z}&=&N\int {\cal D}\tilde{a}^{\mu}\delta(\tilde{a}^{\mu}
\tilde{a}^{\mu}-1)\prod_{i=1}^{2}\delta(\Gamma_i)\det|\{\tilde{\Omega}_i,
\Gamma_j\}|\exp^{i\int {\rm d}t \tilde{L}},  \nonumber \\
\tilde{L}&=&-E+2{\cal I}\dot{\tilde{a}}^\mu\dot{\tilde{a}}^{\mu}  \label{zhct2}
\end{eqnarray}
where $\tilde{L}$ is form invariant Lagrangian of Eq. (\ref{lag}).

Next, in order to derive the exact form of transformation in which the
Lagrangian (\ref{zhct}) is invariant, we use the recently proposed method of
Lagrangian approach \cite{shiz,kkp} which is based on a singular Hessian in
the equations of motion.
Starting with the Lagrangian (\ref{zhct}) with the constraint $\tilde{\Omega}%
_1=a^\mu a^\mu-1+2\theta=0$, we can obtain the equations of motion of the form
\begin{eqnarray}
L^{(0)}_i = W^{(0)}_{ij}\ddot{q}^j + \alpha^{(0)}_i = 0  \label{em}
\end{eqnarray}
where $W^{(0)}_{ij}=\frac{\partial^2 {\cal L}}{\partial \dot{q}^i \dot{q}^j}$
is a Hessian, $\alpha^{(0)}_i=\frac{\partial^2 {\cal L}}{\partial q^j \dot{q}%
^i}\dot{q}^j-\frac{\partial {\cal L}}{\partial q^i}$, and the superscript
for later convenience means the zeroth iteration. If we denote $q^i=(a^\mu,\theta)$,
we have
\begin{eqnarray}
W^{(0)}_{ij}&=& \frac{4{\cal I}}{a^\sigma a^\sigma} \left(
\begin{array}{cc}
\delta_{\mu\nu} & 0 \\
0 & - \frac{1}{a^\sigma a^\sigma}
\end{array}
\right)  \nonumber \\
\alpha^{(0)}_i &=& {\frac{4{\cal I} }{(a^\sigma a^\sigma)^2}} \left(
\begin{array}{c}
- 2 \dot{a}_\mu a^\rho \dot{a}^\rho + a_\mu \dot{a}^\rho \dot{a}^\rho - {%
\frac{2 }{a^\sigma a^\sigma}} a_\mu \dot{\theta}^2 \\
{\frac{4 }{a^\sigma a^\sigma}} a^\rho \dot{a}^\rho \dot\theta
\end{array}
\right)  \label{emc}
\end{eqnarray}
Since the constraint $\Omega_1$ is an A-type defined by a function without
velocities in configuration space, we require as a consistency
condition the following identity
\begin{equation}
L_5 \equiv \frac{{\rm d}^2}{{\rm d}t^2}(\frac{1}{2}\Omega_1 )
= a^\mu \ddot{a}^\mu + \ddot{%
\theta}+ \dot{a}^\mu \dot{a}^\mu =0.  \label{first}
\end{equation}
This requirement is similar to the time stability condition of
constraints in the Hamiltonian formalism. Then, the resulting equation may
be summarized in the form of the set of "first generation" equations
\begin{equation}
L^{(1)}_{i_1} \equiv W^{(1)}_{i_i j}\ddot{q}^j + \alpha^{(1)}_{i_1}
=\left\{
\begin{array}{ll}
L_i^{(0)},\ i=a^\mu,\theta &  \\
\frac{{\rm d}^2}{{\rm d}t^2}(\frac{1}{2}\Omega_1) &
\end{array}
\right.  \label{em1}
\end{equation}
where
\begin{equation}
W^{(1)}_{i_1j}=\left(
\begin{array}{cc}
W^{(0)}_{ij} &  \\ \hline
\begin{array}{cc}
a^\mu & 1
\end{array}
&
\end{array}
\right),~~
\alpha^{(1)}_{i_1}= \left(
\begin{array}{c}
\alpha^{(0)}_i \\
\dot{a}^\mu \dot{a}^\mu
\end{array}
\right).  \label{alp1}
\end{equation}
Since the first iterated Hessian in Eq. (\ref{alp1}) is of rank four, there exists
a null eigenvector satisfying
\begin{equation}
\lambda^{(1)}_{i_1}W^{(1)}_{i_1 j}=0
\end{equation}
from which we can find the solution
\begin{equation}
\lambda^{(1)}_{i_1}=(a^\mu, -a^\mu a^\mu, - \frac{4{\cal I}}{a^\mu a^\mu}).
\label{null}
\end{equation}
In general, the null eigenvectors are known to generate further Lagrange
constraints which is of a function of the coordinates and velocities, but
not of accelerations through $\lambda^{(k)}_{i_k}L^{(k)}_{i_k}=0$ at the
k-th generation of iteration. However, in our case we have
\begin{equation}
\label{sym}
\lambda^{(1)}_{i_1}L^{(1)}_{i_1}=-\frac{2{\cal I}}{(a^\sigma a^\sigma)^2} (%
\frac{{\rm d}}{{\rm d}t}\Omega_1)^2 \approx 0
\end{equation}
which means that no further constraints are generated. The algorithm is then
ended up at this stage.

The symmetries of the Lagrangian (\ref{zhct}) are encoded in the identity
(\ref{sym}), which is a special case of a general theorem \cite{shiz,kkp}
stating that the identity can always be written in the form of
\begin{equation}
\label{form1}
\Omega^{(l)}=\sum_{s=0}^{l} (-1)^s
                \frac{d^s}{dt^s}\phi^{i(s)}_k L^{(0)}_i
             \approx 0
\end{equation}
where the superscript $l$ denotes the last stage of iteration giving the
identity.  The corresponding Lagrangian is then invariant under the transformation
\begin{equation}
\label{form2}
\delta \varphi^i =
\sum_k \left( \epsilon_k \phi^{i(0)}_k
          + \dot\epsilon_k \phi^{i(1)}_k \right).
\end{equation}
In the first class SU(2) Skyrmion, the coefficients $\phi^{i(s)}$
in Eq. (\ref{form1}) are given by
$\phi^{a^\mu (0)}=a^\mu$, $\phi^{\theta (0)}=-a^\mu a^{\mu}$.
As results, by using Eq. (\ref{form2}),
the desired form of symmetry transformation can be read as
\begin{equation}
\delta a^\mu = a^\mu \epsilon,
~~~\delta \theta = - a^\mu a^{\mu}\epsilon.
\label{symm}
\end{equation}
It can be easily checked that the Lagrangian (\ref{zhct}) is invariant
under the transformation (\ref{symm}).

Therefore we have shown that both the Hamiltonian and Lagrangian approaches
have the same symmetry structure because the symmetry transformation rule (\ref{symm})
obtained in the Lagrangian approach is exactly the same as that
in Eq.(\ref{rule}) obtained when we consider
the effective first class constraints (\ref{1stconst})
as the symmetry generators in the pure Hamiltonian formalism.


\section{BRST-BFV analysis for consistent gauge fixing}
\setcounter{equation}{0} \renewcommand{\theequation}{\arabic{section}.%
\arabic{equation}}


Now, in order to obtain the BRST invariant Lagrangian in the framework of the
BFV formalism \cite{bfv,fik,biz} which is applicable to theories with the
first class constraints, we introduce two canonical sets of ghosts and anti-ghosts together with
auxiliary fields $({\cal C}^{i},\bar{{\cal P}}_{i})$, $({\cal P}^{i},\bar{{\cal C}}_{i})$,
$(N^{i},B_{i})$, $(i=1,2)$ which satisfy the super-Poisson algebra
\footnote{
Here the super-Poisson bracket is defined as
\[
\{A,B\}=\frac{\delta A}{\delta q}|_{r}\frac{\delta B}{\delta p}|_{l}
-(-1)^{\eta_{A}\eta_{B}}\frac{\delta B}{\delta q}|_{r}\frac{\delta A} {%
\delta p}|_{l}
\]
where $\eta_{A}$ denotes the number of fermions called ghost number in $A$
and the subscript $r$ and $l$ right and left derivatives.}
\[
\{{\cal C}^{i},\bar{{\cal P}}_{j}\}=\{{\cal P}^{i},\bar{{\cal C}}_{j}\}
=\{N^{i},B_{j}\}=\delta_{j}^{i}.
\]

In the SU(2) Skyrmion model, the nilpotent BRST charge $Q$, the fermionic
gauge fixing function $\Psi$ and the BRST invariant minimal Hamiltonian $%
H_{m}$ are given by
\begin{eqnarray}
Q&=&{\cal C}^{i}\tilde{\Omega}_{i}+{\cal P}^{i}B_{i},  \nonumber \\
\Psi&=&\bar{{\cal C}}_{i}\chi^{i}+\bar{{\cal P}}_{i}N^{i},  \nonumber \\
H_{m}&=&\tilde{H}^{\prime}-\frac{1}{2{\cal I}}{\cal C}^{1}\bar{{\cal P}}_{2}
\end{eqnarray}
which satisfy the relations $\{Q,H_{m}\}=0$, $Q^{2}=\{Q,Q\}=0$, $\{\{\Psi,Q\},Q\}=0$.
The effective quantum Lagrangian is then described as
\begin{equation}
L_{eff}=\pi^{\mu}\dot{a}^{\mu}+\pi_{\theta}\dot{\theta} +B_{2}\dot{N}^{2}+%
\bar{{\cal P}}_{i}\dot{{\cal C}}^{i}+\bar{{\cal C}}_{2} \dot{{\cal P}}%
^{2}-H_{tot}
\end{equation}
with $H_{tot}=H_{m}-\{Q,\Psi\}$. Here $B_{1}\dot{N}^{1} +\bar{{\cal C}}_{1}%
\dot{{\cal P}}^{1}=\{Q,\bar{{\cal C}}_{1} \dot{N}^{1}\}$ terms are
suppressed by replacing $\chi^{1}$ with $\chi^{1} +\dot{N}^{1}$.

Now we choose the unitary gauge
\begin{equation}
\chi^{1}=\Omega_{1},~~~\chi^{2}=\Omega_{2}
\end{equation}
and perform the path integration over the fields $B_{1}$, $N^{1}$, $\bar{%
{\cal C}}_{1}$, ${\cal P}^{1}$, $\bar{{\cal P}}_{1}$ and ${\cal C}^{1}$, by
using the equations of motion, to yield the effective Lagrangian of the form
\begin{eqnarray}
L_{eff}&=&\pi^{\mu}\dot{a}^{\mu}+\pi_{\theta}\dot{\theta} +B\dot{N}+\bar{%
{\cal P}}\dot{{\cal C}}+\bar{{\cal C}}\dot{{\cal P}}  \nonumber \\
& &-E-\frac{1}{8{\cal I}}(\pi^{\mu}-a^{\mu}\pi_{\theta})(\pi^{\mu}-a^{\mu}
\pi_{\theta})\frac{a^{\sigma}a^{\sigma}}{a^{\sigma}a^{\sigma}+2\theta} -%
\frac{1}{4{\cal I}}\pi_{\theta}\tilde{\Omega}_{2}  \nonumber \\
& &+2a^{\mu}a^{\mu}\pi_{\theta}\bar{{\cal C}}{\cal C}+\tilde{\Omega}_{2}N
+B\Omega_{2}+\bar{{\cal P}}{\cal P}
\end{eqnarray}
with redefinitions: $N\equiv N^{2}$, $B\equiv B_{2}$, $\bar{{\cal C}}\equiv
\bar{{\cal C}}_{2}$, ${\cal C}\equiv {\cal C}^{2}$, $\bar{{\cal P}}\equiv
\bar{{\cal P}}_{2}$, ${\cal P}\equiv {\cal P}_{2}$.

Next, using the variations with respect to $\pi^{\mu}$, $\pi_{\theta}$, ${\cal P}$
and $\bar{{\cal P}}$, one obtain the relations
\begin{eqnarray}
\dot{a}^{\mu}&=&\frac{1}{4{\cal I}}(\pi^{\mu}-a^{\mu}\pi_{\theta})
a^{\sigma}a^{\sigma} +a^{\mu}(\frac{1}{4{\cal I}}\pi_{\theta}-N-B)  \nonumber
\\
\dot{\theta}&=&-\frac{1}{4{\cal I}}a^{\mu}(\pi^{\mu}-a^{\mu}\pi_{\theta})
a^{\sigma}a^{\sigma} +a^{\mu}a^{\mu}(-\frac{1}{2{\cal I}}\pi_{\theta}-2\bar{%
{\cal C}}{\cal C}+N) +\frac{1}{4{\cal I}}a^{\mu}\pi^{\mu}  \nonumber \\
{\cal P}&=&-\dot{{\cal C}},~~~~~\bar{{\cal P}}=\dot{\bar{{\cal C}}}
\end{eqnarray}
to yield the effective Lagrangian
\begin{eqnarray}
L_{eff}&=&-E+\frac{2{\cal I}}{a^{\sigma}a^{\sigma}}\dot{a}^{\mu}\dot{a}%
^{\mu} -2{\cal I}\left[\frac{\dot{\theta}}{a^{\sigma}a^{\sigma}} +(B+2\bar{%
{\cal C}}{\cal C})a^{\sigma}a^{\sigma}\right]^{2}  \nonumber \\
& &+\frac{4{\cal I}}{a^{\sigma}a^{\sigma}}a^{\mu}\left[ \dot{a}%
^{\mu}+a^{\mu}(\frac{\dot{\theta}} {a^{\sigma}a^{\sigma}}+(B+2\bar{{\cal C}}%
{\cal C})a^{\sigma}a^{\sigma})\right] (B+N)  \nonumber \\
& &+B\dot{N}+\dot{\bar{{\cal C}}}\dot{{\cal C}}.
\end{eqnarray}

Finally, with the identification
\begin{equation}
N=-B+\frac{\dot{\theta}}{a^{\sigma}a^{\sigma}},
\end{equation}
we obtained the desired BRST invariant Lagrangian of the form
\begin{eqnarray}
L_{eff}&=&-E+\frac{2{\cal I}}{a^{\sigma}a^{\sigma}}\dot{a}^{\mu}\dot{a}%
^{\mu} -\frac{2{\cal I}}{(a^{\sigma}a^{\sigma})^{2}}\dot{\theta}^{2} -2{\cal %
I}(a^{\mu}a^{\mu})^{2}(B+2\bar{{\cal C}}{\cal C})^{2}  \nonumber \\
& &-\frac{\dot{\theta}\dot{B}}{a^{\sigma}a^{\sigma}} +\dot{\bar{{\cal C}}}%
\dot{{\cal C}},
\end{eqnarray}
which is invariant under the BRST transformation
\begin{eqnarray}
\delta_{B}a^{\mu}&=&\lambda a^{\mu}{\cal C},~~~ \delta_{B}\theta=-\lambda
a^{\mu}a^{\mu}{\cal C},  \nonumber \\
\delta_{B}\bar{{\cal C}}&=&-\lambda B,~~~ \delta_{B}{\cal C}=\delta_{B}B=0.
\end{eqnarray}
Here one notes that the above BRST transformation including the rules for
the (anti)ghost fields is just the generalization of the previous one (\ref{symm}%
).

This completes the standard procedure of BRST invariant gauge fixing in the
BFV formalism.


\section{Conclusions}


In summary, we have constructed the first class BFT physical fields, in
terms of which the first class Hamiltonian is formulated to be consistent
with the Hamiltonian with the original fields and auxiliary fields.
After converting the second class SU(2) Skyrmion into the effectively
first class one, we have analyzed the full symmetry structure of the
model purely based on the Hamiltonian approach.
On the other hand, we have also constructed the
effective Lagrangian corresponding to the first class Hamiltonian in the path
integral approach to the partition function.  This Lagrangian includes the new
type of the WZ term restoring the gauge symmetry.
Then, we have
explicitly derived the symmetry structure of this effective Lagrangian through
the Lagrangian approach, showing that the two approaches are equivalent to
give the same symmetry structure of the SU(2) Skyrmion.
Furthermore, in the BFV scheme, we have obtained the BRST
invariant gauge fixed Lagrangian including the (anti)ghost fields, and its
BRST transformation rules.  Finally, through further investigation, the SU(3)
extension \cite{su3}, which is a real phenomenolical model, of this analysis will be
worth to being studied.

\vskip 1.0cm One of us (S.T.H.) would like to thank Professor G.E.
Brown at Stony Brook for constant concerns and encouragement.  The
present work was partly supported by the BK21 Project No. D-0055,
Ministry of Education, 1999 and the Sogang University Research
Grants in 1999.


\end{document}